# *From Requirements to Java code: an Architecture-centric Approach for producing quality systems*

*Antonio Bucchiarone*, IMT of Lucca, Piazza San Ponziano, 6. 55100 Lucca, Italy, a.bucchiarone@imtlucca.it

*Davide Di Ruscio*, University of L'Aquila, Computer Science Department, Via Vetoio, 1. 67100 L'Aquila, Italy. diruscio@di.univaq.it

*Henry Muccini*, University of L'Aquila, Computer Science Department, Via Vetoio, 1. 67100 L'Aquila, Italy. muccini@di.univaq.it

*Patrizio Pelliccione*, University of L'Aquila, Computer Science Department, Via Vetoio, 1. 67100 L'Aquila, Italy. pellicci@di.univaq.it

**Abstract**. *When engineering complex and distributed software and hardware systems (increasingly used in many sectors, such as manufacturing, aerospace, transportation, communication, energy, and health-care), quality has become a big issue, since failures can have economics consequences and can also endanger human life. Model-based specifications of a component-based system permit to explicitly model the structure and behaviour of components and their integration. In particular Software Architectures (SA) has been advocated as an effective means to produce quality systems. In this chapter by combining different technologies and tools for analysis and development, we propose an architecture-centric model-driven approach to validate required properties and to generate the system code. Functional requirements are elicited and used for identifying expected properties the architecture shall express. The architectural compliance to the properties is formally demonstrated, and the produced architectural model is used to automatically generate the Java code. Suitable transformations assure that the code is conforming to both structural and behavioural SA constraints. This chapter describes the process and discusses how some existing tools and languages can be exploited to support the approach.*

## 1. Introduction

The typical use of SA is as a high level design blueprint of the system to be used during the system development and later on for maintenance and reuse (in order to capture and model architectural design alternatives). At the same time, SA can be used in order to analyze and validate architectural choices, both behavioural and quantitative (by complementing traditional code-level analysis techniques). More recently, architectural artefacts have been used to implicitly or explicitly guide the design and coding process (ARCHJAVA Project, 2005; Fujaba Project, 2006). In summary, SA specifications are nowadays used for many purposes (Mustapic, 2004; Bril, 2005; Bernardo, 2003) like documenting, analysing, or to guide the design and coding process.

Even though SA documentation, analysis, and code generation have been intensively analyzed in isolation (e.g., Bernardo, 2003; Muccini, 2006; Fujaba Project, 2006) (code generation only very recently and partially), a tool supported process for selecting and documenting the right architecture and for successively propagating architectural design to the final system implementation is still missing.

Analysis techniques and tools have been introduced to understand if the SA satisfies certain expected properties. By using Model Checking, testing, performance analysis (and others) at the architectural level, a software architect can assess the architectural quality and predict the final system characteristics. In the context of code generation, this verification phase assumes even a more central role, being the selected architectural model used for (automatically) deriving the system implementation. However, most of the analysis techniques rely on formal architectural specifications (e.g., (Bernardo, 2003; Muccini, 2006)) of difficult application in industrial projects and of difficult integration in the software development process.

In this chapter we propose an architecture-centric development approach which enables the Java code generation of a software system from a high quality architectural model-based design. High quality architecture hereafter is referred to the SA ability to fulfill certain functional and temporal constraints as imposed by the requirements. Other qualities are not explicitly taken into consideration. The formally verified SA is then the starting point of model transformations that produce a skeleton of the Java code of the system. The produced code reflects both structural and behavioural SA constraints and consequently assures the validity of defined and verified functional requirements.

Thus the goals are twofold: *to validate the model-based architectural specification with respect to defined requirements, and to use this validated model to guide the generation of a quality system implementation using a model-driven approach.*

*Key benefits* of this approach are manifold: a model-based specification of the SA is provided, the conformance relation between functional requirements and architecture is validated, Java code is automatically generated from architectural models. The generated code is obliged to respect both structural SA constraints (i.e., each component can only communicate using connectors and ownership domains that are explicitly declared in the SA) and behavioural constraints (i.e., methods provided by components can be invoked only consistently to the behaviours defined for the components). The approach is supported by automated tools, which allow formal analysis and permits code generation from the validated architecture. Overall, the approach encourages developers to make a more extensive and practical usage of SA specifications.

The following of the chapter is organized as follows: Section 2 provides state of the art information on functional requirements specification, on SA modelling and analysis and on code generation. Based on this background information, Section 3 will describe our proposal for an architecture-centric model-driven and quality oriented development process from requirements to code.. Section 4 introduces an ATM system running case study that is used for detailing the approach. Section 5 outlines future research directions while Section 6 concludes the chapter.

## *2. Background*

This section provides background information on the state of the art on functional requirements specification (Section 2.1), on formal and model-based specification of SAs (Section 2.2), on architecture-level analysis (Section 2.3), and on existing code generation techniques from architectural specifications (Section 2.4).

### 2.1 Functional Requirements Specification

Some work has been proposed in the last years attempting to bridge the gap between an informal functional requirements description to a formal one. Works in this area, related to our proposed research, can be organized into three groups: *properties elicitation and formalization*, *approaches to bring the gap between informal requirements' descriptions and formal ones*, and *requirements to architecture transition*.

**Properties**: In the literature little attention has been put in the properties to be proven. In general, these are assumed to exist as part of the problem specification. Holzman in (Holzmann, 2002) states that one of the "most underestimated problem in applications of automated tools to software verification" is "the problem of accurately capturing the correctness requirements" (properties) "that have to be verified" and continues identifying the difficulty of such task. When the verification technique is model checking (Clarke, 2000), temporal logic is the standard method to express the correctness requirements. In the same chapter, Holzman shows how Linear-time Temporal Logic (LTL) (Manna, 1992) formulae may be used to describe properties and how the level of sophistication required by them may allow one to specify properties in a wrong way. However, in an industrial context it is unfeasible to write by hand complex LTL formulae. To this extent, he proposes a tool to write temporal properties in a graphical notation.

In (Smith, 2002) the authors recognize the difficulty in writing properties correctly. They notice that this difficulty is not only related to the chosen notation: "no matter what notation is used, however, there are often subtle, but important, details that need to be considered". In order to mitigate this problem, they propose PROPEL introducing pattern templates previously identified, which are represented using both disciplined natural language and finite state automata.

**From Informal to Formal Requirements specification**: Many languages and notations have been suggested and devised for use in requirements engineering. Less formal notations, such as scenarios and use cases, have proven to be more effective for elicitation, negotiation, and validation phases, while more formal notations have proven more effective for requirements specification and analysis. Much work has been proposed in the last years attempting to bridge the gap between informal requirement descriptions to formal ones. We here discuss only those works we believe closer to our approach.

Scenify (Hähnle, 2002) is a natural-language processing tool that translates natural-language (English) input into a schematic representation.

Johannisson in its PhD thesis (Johannisson, 2005) investigates how to bridge the gap between formal and informal software specifications. This work makes use of interactive syntax-directed editor, parsers and linearizers, based on a grammatical framework that combines linguistic and logical methods. The approach proposed in this chapter is related to a number of other approaches that have been considered by researchers.

In (Zhu, 2003) authors exploit a software tool that allows system engineers to write detailed use case descriptions using structured templates. The specification is guided by use case style guidelines, temporal semantics and an extensive dictionary of naval domain nouns. Once the use case description phase has been accomplished, system engineers derive use case specifications and, after parameterization, corresponding scenarios are automatically generated.

In the Specification Pattern Instantiation and Derivation EnviRonment (SPIDER)framework (Cheng, 2006), developers can create natural language specifications of properties that are automatically and transparently mapped to the property specification language of the targeted analysis tools, e.g., LTL.

**From Requirements to Software Architecture**: The problem of deciding how requirements, architectures and implementation have to be mutually related is still open as advocated and investigated by many researchers (STRAW 2003; Nuseibeh, 2001; Grünbacher, 2003). In (STRAW 2003; Nuseibeh, 2001), ways to bridge the gap between requirements and SAs have been proposed. Egyed proposes ways to trace requirements to SA models (Grünbacher, 2003) and SA to the implementation (Medvidovic, 2003).

**Considerations:** One relevant problem that arises during the requirement engineering process is the result of failing to make a clear transition between different levels of requirements description. According to the terminology adopted in (Sommerville, 2004), the term "user requirements" is used to mean high-level abstract requirement descriptions and the term "system requirements" is used to mean detailed and possibly formal descriptions. Often in practice, stake-holders are able to describe user requirements in an informal way without detailed technical knowledge. They are rarely willing to use structured notations or formal ones. Transiting from user requirements to system requirements is an expensive task. In fact, we are speaking about decisions made during this early phase of the software development process, when the system under development is vague also in the mind of the customer. A good answer to this need is W_PSC (Autili, 2006), a speculative tool that facilitates understanding and structuring requirements. By means of a set of sentences (based on expertise in requirements formalization and on a set of well-known patterns (Dwyer, 1999) for specifying temporal properties used in practice) and classified according to main keywords of temporal properties, W_PSC forces to make decisions that break the uncertainty and the ambiguity of user requirements.

The output of W_PSC is a temporal property expressed in Property Sequence Chart (PSC) (Autili, 2007). PSC is a simple and (sufficiently) powerful formalism for specifying temporal properties in a user-friendly fashion. It is a scenario-based visual language that is an extended graphical notation of a subset of UML2.0 Sequence Diagrams. PSC can graphically express a useful set of both liveness and safety properties in terms of messages exchanged among the components forming the system. W_PSC supports also the user on taking many decisions required transiting from requirements to architecture. Indeed, automatically transforming informal requirements into formal temporal properties is not always possible (due to inconsistencies or under specifications) and may become time consuming. W_PSC, as all those related approaches previously summarized, makes an attempt to make the transition from informal requirements to formal properties easier and faster. Details about W_PSC and PSC will be provided on the following Sections 3 and 4.

## 2.2 Software Architecture Specification

Two main classes of languages have been used so far to specify SAs: formal Architecture Description Languages (ADLs) and model-based specifications with UML.

Many ADLs have been proposed in the last fifteen years, with different requirements and notations, and with the objective to support components' and connectors' specification and their overall interconnection, composition, abstraction, reusability, configuration, heterogeneity, and analysis mechanisms (Medvidovic, 2000). Table 1 shows the most known ADLs evidencing the ones still supported. The table contains also approaches which are usually classified as non-conventional ADLs since they possibly neglect fundamental aspects.

Even if much work has been done on this direction, the application of such techniques into industrial systems can still be very difficult due to some extra requirements and constraints imposed by realistic scenarios (Bertolino, 2004): as a consequence, we cannot always assume that formal modelling of the software system exists. On the contrary, a semi-formal, easy to learn and possibly diagrammatic notation may reasonably offer enough pragmatic qualities. When software architects start defining the SA of system requirements, needs and challenges of the system are not well established but significant decisions have to be made. Significant decisions at the SA level encompass the organization of a software system, the selection of the structural elements, their interfaces, their behaviours, and the composition of these elements into progressively larger subsystems. More often than not all these aspects are not very well defined and cannot be established one forever, but the architect must move forward accepting ambiguities. In practice the software architect must made decisions and have to choose a solution in a partly dark and then is obliged to select a solution that probably is a suboptimal solution (Hofmeister, 2007). Ideas and constraints coming from different stakeholders (such as end-users, customers, developers, sales and field support, maintainers, development managers, system administrators) together with architectural requirements constitute the nebulous set of architectural constraints and requirements. Walking in partly dark the software architect selects a first version of the SA.

Summarizing, the application of modelling and analysis techniques for the use of SAs in practice imposes some extra requirements: the methodology must be tool supported and in general it is unacceptable the use of tools that require a great effort and require spending a lot of time for reaching the point from where the tool become useful.

With the introduction of UML as the de-facto standard to model software systems and with its widespread adoption in industrial contexts, many extensions and profiles have been proposed to adapt UML to model architectures. Many proposals have been presented so far to adapt UML 1.x to model SAs (e.g., (Robbins, 1998; Kruchten, 1995; Gomaa, 2001; Kande', 2002)). In such proposals, researchers have compared the architectural needs with UML concepts, extended or adapted UML, or created new profiles to specify architecture specific needs with UML.

| ADL | Born Data | Tools | Still Supported | Notes |
|---|---|---|---|---|
| Rapide | 1990 | Rapide | NO | ADL and simulation |
| Darwin | 1991 | LTSA + SAA | YES | Focus on dynamic SA |
| Weaves | 1991 | Weaves | NO | Data-flow-architectures with high-volume of data |
| Adage | 1992 | — | NO | Avionics navigation and guidance Architecture Description |
| LILEANNA | 1993 | LILEANNA | NO | Modules connection language |
| MetaH & MetaS | 1993 | MetaH | YES | ADL for avionic domain |
| ArTek | 1994 | — | NO | Non conventional ADL |
| Resolve | 1994 | Resolve | NO | Focus on Component Specification |
| Wright | 1994 | Wright | NO | Focus on communications |
| Acme | 1995 | AcmeStudio | YES | Interchange Language |

|  |  | Armani |  | between ADLs |
|---|---|---|---|---|
| SADL | 1995 | Sadl tool | NO | Focus on Refinement |
| UniCon | 1995 | UniCon | NO | Focus on connectors and Styles |
| C2 SADEL & C2 AML | 1996 | Dradel, SAAGE ArchStudio | NO | ADL based on C2 style |
| GenVoca | 1996 | P3 | NO | Non conventional ADL |
| Fujaba | 1997 | Fujaba | YES | Non conventional ADL |
| Jacal | 1997 | Jacal 2 | YES | Focus on prototyping SA |
| Koala | 1997 | Koala tools | YES | ADL for product families |
| Little-JIL | 1998 | Little-JIL 1.0 | NO | Non conventional ADL |
| Maude | 1998 | Maude 2.0 | YES | Non conventional ADL |
| ADML | 2000 | ADML Enabled Tools | YES | XML-based ADL |
| xArch/xADL | 2000 | xADL 2.0 | YES | XML-based ADL |
| AADL | 2001 | Osate | YES | Embedded real-time systems / Avionics systems |
| xArch/xAcme | 2001 | AcmeStudio | YES | Acme in XML |
| ABC/ADL | 2002 | ABC tool (prototype) | YES | ADL for component composition |
| Prisma | 2002 | PrismaCase | YES | Component-based systems |
| DAOP-ADL | 2003 | DAOP-ADTools | YES | Component and Aspect-based ADL |

**Table 1:** The most known ADLs

Recently, several works propose UML 2.0 native specifications (i.e., without any profile or extension) for SA modelling. In (Eriksson, 2004) logical architectures, patterns and physical architectures are represented by using components, dependencies, and collaborations. In (Pender, 2003) components within a component diagram are used to model the logical and physical architecture. In order to bridge the gap between UML 2.0 and ADLs, some aspects still require adjustments, thus much work is still ongoing (Goulo, 2003; Roh, 2004; Ivers, 2004; Perez-Martinez, 2004).

The success of model-based specifications of SAs is proven by many profiles defined so far for UML-modelling of SAs (e.g., (AADL; SysML)). Unfortunately, UML does not have a well defined semantics and then is open to ambiguities and misunderstanding. Moreover, while such UML notations are well suited to model some aspects of SAs, they are agnostic of others (Dashofy, 2002; Medvidovic, 2002).

**2.3 Software Architecture Analysis**

While how to model SAs has been for a long time the main issue in the SA community, how to select the right architecture has become one of the most relevant challenges in recent days. Model Checking, deadlock detection, testing, performance analysis, and security are, among others, the most investigated analysis techniques at the architectural level. Among the techniques that allow designers to perform exhaustive verification of the systems (such as theorem provers, term rewriting systems and proof checkers) model checking (Clarke, 2000) has as main advantage that it is completely automatic. The user provides a model of the system and a specification of the property to be checked on the system and the model checker provides either true, if the property is verified, or a counter example is always generated, if the property is not valid. The counter example is particularly important since is show a trace that leads the system to the error condition.
While presenting a comprehensive analysis of the state of the art in architectural analysis is out of the scope of this chapter, this section will focus on architecture-level Model Checking techniques. For further reading on the topic, interested readers may refer to e.g., (Bernardo, 2003; Muccini, 2006; Dobrica, 2002).

Initial approaches for Model Checking at the architecture level have been provided by the Wright architectural language (Allen, 1997) and the Tracta approach (Magee, 1999). More recently, many other approaches have been proposed, as listed and classified in Figure 1. By focussing on the model-based approaches, Bose (Bose, 1999) presents a method which automatically translates UML models of SA for verification and simulation using SPIN (Holzmann, 2003). A component is specified in terms of port behaviours and performs the computation or provides services. A mediator component is specified in terms of roles and coordination policies. Safety properties are checked. Lfp (Jerad, 2005) is a formal language dedicated to the description of distributed embedded systems' control structure. It has characteristics of both ADL and coordination language. Its model checker engine is Maude based on rewriting logic semantics. Fujaba (Fujaba Project, 2006) is an approach tool supported for real-time Model Checking of component-based systems: the system structure is modelled through UML component diagrams, the real-time behaviour is modelled by means of real-time statecharts (an extension to UML state diagrams), properties are specified in TCTL (Timed Computation Tree Logic) (Alur, 1990) and the UPPAAL (UPPsala and AALborg University) (Bengtsson, 1995) model checker is used as the real-time model checker engine. Arcade (Barber, 2001) (Architecture Analysis Dynamic Environment) applies model checking to a DRA (Domain Reference Architecture) to provide analysts and developers with early feedback from safety and liveness evaluations during requirements management. The properties are represented as LTL formulae and the model checker engine is Spin. AutoFOCUS (AutoFOCUS Project) is a model-based tool for the development of reliable embedded systems. In AutoFOCUS, static and dynamic aspects of the system are modelled in four different views: structural view, interaction view, behavioural view, and data view. AutoFOCUS provides an integrated tool for modelling, simulation, and validation. AutoFOCUS2 (AutoFOCUS2 Project) advances and improves previous work on AutoFOCUS by adding new modeling views.

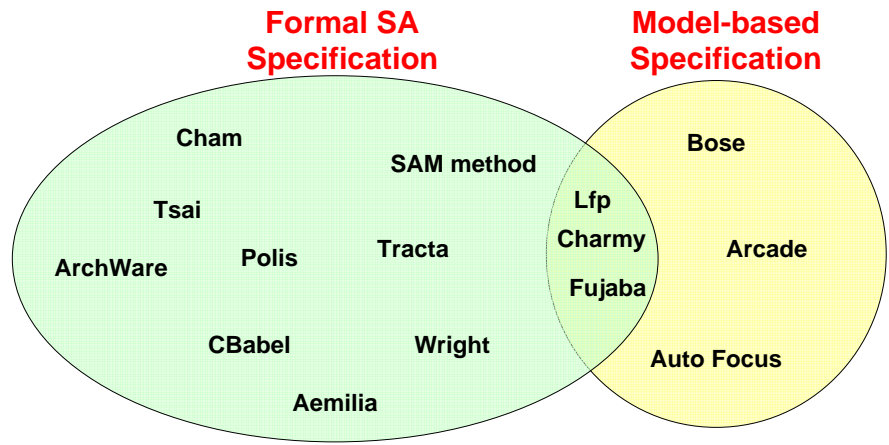

**Figure 1:** Model Checking Techniques based on Formal or Model-based Architectural Specifications

CHARMY (Pelliccione, 2005; CHARMY, 2004; Inverardi, 2005) is our proposal to model-check SA compliance to desired functional temporal properties. It intends to fill this gap by providing an automated, easy to use tool for the model-based design and validation of SA. CHARMY main strengths are as follow:

**Informal vs formal**: formal languages allows for automatic analysis, but they are generally time and cost consuming, while requiring certain specific skills. Informal languages, instead, are faster and easier to learn, by permitting lower automation. CHARMY tries to incorporate both advantages, and mitigate their respective weaknesses automatically completing informal and incomplete models: SA topology and behaviour are described via UML based specifications and automatically translated into a formal prototype. In particular, components and state diagrams, used to specify the SA topology and behaviour, are automatically interpreted to synthesize a formal Promela prototype, which is the SPIN model checker modelling language;

**SA simulation and checking**: CHARMY provides support for simulating the SA: it uses the SPIN simulation engine and offers simulation features which interpret SPIN results in terms of CHARMY state machines. Moreover, properties whose validity needs to be checked on the architectural model are modelled through scenarios, by expressing desired and undesired behaviours. Such scenarios are automatically translated into Büchi automata (Büchi, 1960), an operational representation for LTL formulae. SPIN is then used to check the conformance of the Promela prototype with respect to such behavioural properties;

**Automatic tool support**: the CHARMY approach for specifying and analyzing SAs is tool supported, and it hides most of the complexity of the modelling and analysis process. Model-based architectural specifications, drawn using the CHARMY editors or standard UML tools, are automatically translated into a formal prototype. The prototype can then be automatically checked for correctness with respect to desired properties.

More details will be provided in Section 4 when we will present the running case study.

### 2.4 Code Generation from Software Architecture Specification

In this section we present an overview on the various techniques used to generate code from a SA specification. We focus the attention on languages that can be used to generate code from a high-level description of the SA. They can be distinguished in Architecture Description Languages (ADLs), such as languages for describing SAs, and Architectural Programming Language (APL), such as languages that integrate SA concepts into programming languages. We conclude this section with a comparison among APLs. It is important to note that code generated from ADLs not necessarily contain architecture concepts. This can have impact on the readability of the code and can reduce its modifiability and maintainability. Furthermore, modifications on the generated code made by developers can invalidate architectural constraints. APLs have been introduced to solve this problem. All these aspects will be detailed in the following.

### 2.4.1 ADLs and code generation

Some ADLs support code generation from an architectural description of the system. In Table 2 we list the ADLs that support code generation: it shows the ADL name, the tool support and the type of code that they produce as output. We have considered ADLs that are currently used in an industrial context and that are continuously updated showing the last release and the references.

| **ADL** | **Data Born** | **Tool Support** | **Output Code** | **Last Update** | **Reference** |
|---|---|---|---|---|---|
| Darwin | 1991 | LTSA-WS + SAA | C++ | March 2007 | (Magee, 1999) |
| Fujaba | 1997 | Fujaba | Java | July 2007 | (Fujaba Project, 2006) |
| xArch/xADL | 2000 | ArchStudio + Apigen | XML | January 2005 | (xADL 2.0, 2005) |
| AADL | 2001 | Osate | Ada, C, Java | April 2007 | (AADL) |
| Prisma | 2002 | PrismaCase | C# | September 2007 | (PRISMA) |

**Table 2:** Code Generation from ADLs

However, the implementation step is, at best, only supported by code generation facilities not capable of explicitly representing architectural notions at the code level. Thus, the notion of SA components, connectors and configurations is kept implicit and the implementation inevitably tends to loose its connection to the intended architectural structure during the maintenance steps. The result is "architectural erosion" (Perry, 1992).

### 2.4.2 Architectural Programming Languages (APLs)

APLs overcome the problem of architectural erosion in implementations by integrating SA concepts into programming languages.

With APLs, there is an inclusion of architectural notions, like components, ports with provided and required interfaces as well as protocols, connectors, and assemblies, into a programming language (typically Java). The basic idea of architectural programming is to preserve the SA structure and properties throughout the software development process so as to guarantee that each component in the implementation may only communicate directly with the components to which it is connected in the architecture. In fact our objective is to have a development process that guarantees the "Communication Integrity" between code and SA (xADL 2.0, 2005).

In this section we present ARCHJAVA and JAVA/A, which are the most famous and advanced APLs (Baumeister, 2006), in order to understand their main characteristics and to compare them with respect to aspects that we consider important for an APL. Then, based on the proposed comparison, we will choose one of the two technologies to be part of our SA-based quality process.

### ARCHJAVA

ARCHJAVA (ARCHJAVA Project, 2005) is an APL which extends the Java language with component classes (which describe objects that are part of the architecture), connections (which enable components communication), and ports (which are the endpoints of connections).

Components are organized into a hierarchy using ownership domains, which can be shared along connections, permitting the connected components to communicate through shared data. A component in ARCHJAVA is a special kind of object whose communication patterns are explicitly declared using architectural declarations. Component code is defined in ARCHJAVA using *component classes*. Components communicate through explicitly declared ports. A *port* is a communication endpoint declared by a component. Each port declares a set of required and provided methods. A provided method is implemented by the component and is available to be called by other components connected to this port. Conversely, each required method is provided by some other component connected to this port. Each provided method must be implemented inside the component.

ARCHJAVA requires developers to declare the connection patterns that are permitted at run time. Once connect patterns have been declared, concrete connections can be made between components. All connected components must be part of an ownership domain declared by the component making the connection.

*Communication integrity* is the key property enforced by ARCHJAVA, ensuring that components can only communicate using connections and ownership domains that are explicitly declared in the architecture. ARCHJAVA guarantees communication integrity between an architecture and its implementation, even in the presence of advanced architectural features like run time component creation and connection. A prototype compiler for ARCHJAVA is publicly available for download at the ARCHJAVA web site (ARCHJAVA Project, 2005).

A. Bucchiarone, D. Di Ruscio, H. Muccini, P. Pelliccione, *From Requirements to Java code: an Architecture-centric Approach for producing quality systems*

***Example in* ARCHJAVA**

We illustrate ARCHJAVA through a simple example. Figure 2 shows a UML composite component diagram of the toy example[1]

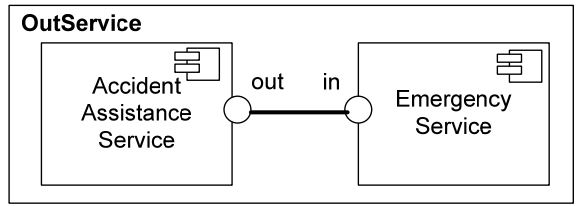

**Figure 2:** OutService Composite Component in ARCHJAVA

```
public component class outService {
protected owned AccidentAssistanceService aas = …;
protected owned EmergencyService es = …;

connect pattern AccidentAssistanceService.out,
        EmergencyService.in;

public outService() {
        connect (aas.out, es.in);
}
}

public component class EmergencyService{
public port in{
        provides void AlertEmergencyService(int loc);
        provides void EmergencyLevel(int level);
        provides void AlertAccepted();
}
public void AlertEmergencyService(int loc){
        …
}
public void EmergencyLevel(int level){
        …
}
public void AlertAccepted(){
        …
}
 }

public component class AccidentAssistanceService{
public port out{
        requires void AlertEmergencyService(int loc);
        requires void EmergencyLevel(int level);
        requires void AlertAccepted();
            }
}
```

---

[1] Borrowed from Sensoria, Research supported by the EU within the FET-GC2 IST-2005-16004 Integrated Project Sensoria (Software Engineering for Service-Oriented Overlay Computers)

The *OutService* component is made up of two subcomponents: the *AccidentAssistanceService* (AAS) and the *EmergencyService* (ES). The first has one *out* port and the second one an *in* port through which the two components are connected. A port is a communication endpoint declared by a component. For each port the language provides constructs to define *requires* and *provides* methods. ARCHJAVA requires developers to declare in the architecture the connection patterns that are permitted at run time.

Taking a look to the code for the specification in Figure 2, the declaration "connect pattern" in our code permits the *OutService* component to make connections between the *out* port of its AAS subcomponents and the *in* port of its ES subcomponent. Once a connect pattern has been declared, concrete connections can be made between components. For example the *constructor* for *OutService* connects the *out* port of the AAS component instance to the *in* port of the ES component instance. This connection binds the required methods (*AlertAccepted*, *AlertEmergencyService*, etc.) in the *out* port of AAS to a provided method with the same name and signature in the *in* port of ES component. Thus when AAS invokes *AlertAccepted* on its *out* port, the corresponding implementation in ES will be invoked.

*JAVA/A*
The basic idea of JAVA/A (Baumeister, 2006; Hacklinger, 2004) (as in ARCHJAVA) is to integrate architectural concepts, such as components, ports and connectors, as fundamental parts into Java. The underlying component model is compatible with the UML component model (Hacklinger, 2004; OMG). This compatibility and the one-to-one mapping of these concepts allow software designers to easily implement UML 2.0 component diagrams. They can express the notions present in these diagrams using built-in language concepts constructs of Java. Furthermore, the visibility of architectural elements in the JAVA/A source code prevents architectural erosion.

The basic concepts of the JAVA/A component model are *components*, *ports*, *connectors* and *configurations*. Any communication between JAVA/A components is performed by sending messages to ports. This message must be an element of the required interface of the perspective port. The port will then pass on the message to the attached connector, which itself will delegate the message to the port at its other end. Each port may contain a *protocol*. These protocols describe the order of messages that are allowed to be sent from and to the respective port. Any incoming and outgoing communication must conform to the protocol. Protocols are realised by UML state machines and ensure the soundness of a configuration at compile-time. A *Connector* in JAVA/A links two components by connecting ports they own.

The JAVA/A compiler is not yet complete and available but authors claim that it will transform JAVA/A components into pure Java code which can be compiled to byte code using the Java compiler. It will be possible to compile and deploy each component on its own, since the component's dependencies on the environment are encapsulated in ports. The correctness of an assembly (i.e., deadlock-freedom) can be ensured using the UML state machine model checker HUGO (HUGO, 2005). Another important aspect that JAVA/A has is the *dynamic reconfiguration*. It summarises changes to a component-based systems at runtime, concerning creation and destruction of components and building up and removing connections between ports. JAVA/A supports each of these reconfiguration variants. JAVA/A has a semantic model that uses a states as algebras approach (Baumeister, 2006) for representing the internals of components and assemblies, and the I/O-transition systems for describing the observable behaviour.

*Example of JAVA/A*
Figure 3 shows a composite component diagram of the same system already introduced for ARCHJAVA (in Fig. 2).

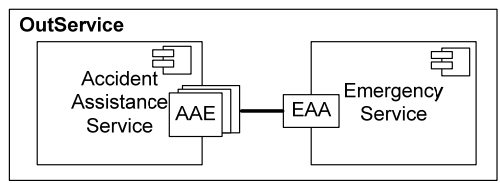

**Figure 3:** OutService Composite Component in JAVA/A

The composite component contains an assembly of two components Accident Assistance Service (AAS) and Emergency Service (ES) whose ports are wired by a connector. The AAE port of the AAS component is depicted as stacked boxes since it is a dynamic port which can have an arbitrary number of port instances. In contrast, the static port EAA must have a single instance at any time. Port protocols are specified with UML state machines. A protocol describes the order and dependencies of messages which are sent and received by a port. The code corresponding to this specification is described and illustrated in Appendix A.

*A Comparison*
ARCHJAVA and JAVA/A employ similar approaches. Both augment java with the concepts of component and connector. ARCHJAVA components have ports with required and provided interfaces. However, ports in ARCHJAVA do not have associated protocols. As a result the dynamic behaviour of ports is not captured in ARCHJAVA.

ARCHJAVA, as well as JAVA/A, allows hierarchical component composition. In JAVA/A there is no possibility of communicating with components other than sending messages to their ports, whereas in ARCHJAVA outer components can invoke methods of inner components directly, which breaks the encapsulation. While ARCHJAVA lacks a semantic model, JAVA/A provides a complete one based on algebras and I/O- transitions systems. As far as concern tool support, in (Schmerl, 2004) the authors have developed additional Eclipse plug-ins that integrates AcmeStudio (Acme) and ARCHJAVA. With this framework an architect can model an architecture using AcmeStudio, and have access to AcmeStudio's verification engines to check desired architectural properties. The architect can then generate ARCHJAVA code using the refinement plug-in. As developers complete the implementation to provide the functionality of the system, ARCHJAVA's checks help ensuring that the implementation conforms to the architect's design. Unfortunately the existing ARCHJAVA environment supports only the *verification* of architectural properties and it does not *force* the developers to respect the component behaviour described into the SA. For JAVA/A the tool support is not yet complete and it is one of future work. So far, a JAVA/A compiler should transform JAVA/A components into pure Java code which can be compiled to byte code using the Java compiler. However, this compiler is not yet publicly available.

Table 3 synthesizes the above discussion and way of understanding the key features and differences of ARCHJAVA and JAVA/A.

| APL | Components | Ports | Configurations | Encapsulation |
|---|---|---|---|---|
| ARCHJAVA | Yes | Yes | Implicit | Partial |
| JAVA/A | Yes | Yes | Explicit | Yes |

| APL | Behavioural Modeling | Distributed Applications | Asynchronous Communication | Tool Support |
|---|---|---|---|---|
| ARCHJAVA | No | No | No | Total |
| JAVA/A | Yes | Yes | Explicit | Not yet |

**Table 3:** APLs Comparison

Our SA-based approach makes use of the ARCHJAVA language since the availability of the corresponding compiler has allowed us to develop each phase of the approach described in the next Section leading to a prototypical implementation available for download at (CHARMY, 2004).

## *3. The proposed Approach*

The model-based architecture-centric and automated analysis approach we are proposing aims at combining exhaustive analysis techniques (Model Checking) and SA-based code generation to produce highly-dependable systems in a model-based development process. Figure 4 shows the activities of the architecture-centric analysis and deployment approach. It is composed of four principal activities: *(i)* specification of functional requirements, *(ii)* model-based specification of SAs, *(ii)* validation of the SA specification with respect to requirements through Model Checking, and finally *(iv)* architecture-based code generation.

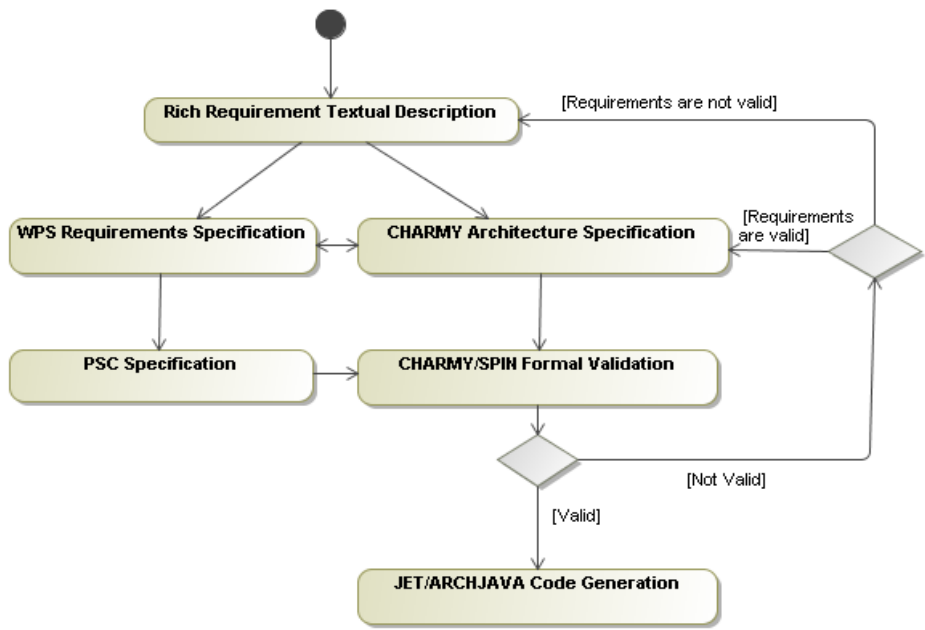

**Figure 4:** The proposed approach

The architectural topology and behaviour is captured through a UML-based notation, part of the CHARMY framework (Pelliccione, 2005; Inverardi, 2005). Properties are elicited from requirements by means of W_PSC (Autili, 2006) and modelled according to the Property Sequence Chart (PSC) (Autili, 2007) language. CHARMY is used to check the architectural model conformance with respect to identified functional properties. After this activity, the architecture is proven to be compliant with selected properties. In the SA-based code generation activity CHARMY models are translated into ARCHJAVA code by means of a developed code generator based on the Eclipse Java Emitter Template (JET) framework (part of the Eclipse Modeling Framework (Budinsky, 2003)). Finally, the ARCHJAVA compiler is used to generate Java code and to ensure that the implementation conforms to the architectural specification.

In the following, by referring to Figure 4, each activity of the proposed approach is individually described.

## 3.1 W_PSC Requirements Specification and formalization in the PSC language

Functional requirements are identified, modeled and analyzed. In order to automatically verify that the system SA satisfies the functional requirements they are typically expressed and formalized as formulae in temporal logics. Unfortunately, the level of inherent sophistication required by these formalisms too often represents an impediment to move these techniques from "research theory" to "industry practice". PSC facilitates the non trivial and error prone task of specifying, correctly and without expertise in temporal logic, temporal properties. PSC can graphically express a useful set of both *liveness* and *safety* properties in terms of messages exchanged among the components forming the system. Finally, an algorithm, called PSC2BA (Autili, 2007), translates PSC into a temporal property representation understandable by model checkers.

Since the aim of this chapter is not to present PSC (presented elsewhere) we do not provide details about this language, but we refer to (Autili, 2007) for a fully description of both the textual and graphical language and its declarative and operational semantics. Moreover, in Section 4 we explain further aspects of the language as needed to fully understand the approach. While the translation process from PSC diagram to Büchi automata is fully automated, the selection of properties from requirements and their formalization in PSC is totally left to engineers' experience. Both tasks may become expensive and error prone, when applied to real projects. W_PSC aims at alleviating the engineers work in eliciting and formalizing properties bridging the gap between possibly informal requirement specifications (as found in practice) and formal ones (as needed in formal methods). It is a conversational tool that, by means of well structured and deep sentences, helps the software engineers in identify and formalize properties. It has been built selecting and classifying the PSC statements, and permits to incrementally build PSC diagrams, starting from user requirements. W_PSC offers a user-friendly wizard helpful while translating a user requirements description into PSC scenarios. It is composed of several windows that present sentences helpful for requirements understanding and selection. The sentences are grouped according a classification based on temporal properties keywords. Since PSC is built by taking into account the same keywords, W_PSC introduces an intuitive way to use all the subtle and precise instruments of PSC. In Section 4 we will provide further details of W_PSC as needed for understanding the case study, while we refer to (Autili, 2006) for a complete description of W_PSC.

### 3.2 Software Architecture model-based Specification in CHARMY

The SA is designed in CHARMY (CHARMY, 2004; Inverardi, 2005) that allows software engineers to specify both the structure and the behaviour by using UML-based notations. We use CHARMY to design the SA since it provides automatisms for verifying the SA by means of model checking techniques. CHARMY allows the specification of the SA topology in terms of components and relationships among them, where components represent abstract computational subsystems. As shown in Figure 5, the internal behaviour of each component is specified in terms of CHARMY state diagrams. The CHARMY notation for state machines permits to specify the intra-component and inter-component behaviours of architectural components and connectors (i.e., the internal behaviour of architectural elements and their integration, respectively). States of the state machines are connected by means of transitions. Transitions are labelled by transitions name and could represent either a message sent or received, denoted by an exclamation mark "!" or a question mark "?", respectively.

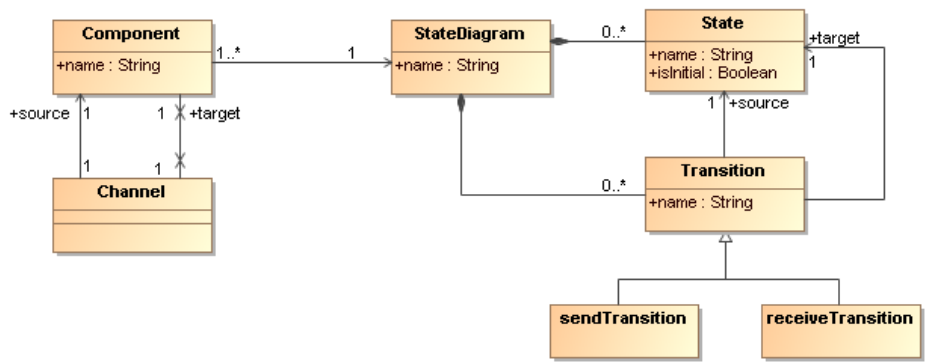

**Figure 5:** Chunks of the CHARMY Metamodel

In Figure 7 a sample CHARMY specification that will be considered throughout the section is depicted. In particular, it is a model that conforms to the metamodel in Figure 5 and it consists of the components *C1* and *C2* connected through the channels *C1_TO_C2* and *C2_TO_C1*. For each component, a corresponding state machine is provided in order to describe the admitted component behaviours.

### 3.3 Software Architecture Verification

CHARMY uses Model Checking techniques to validate the SA conformance to certain properties. Being the SA typically used as the driver for the entire development process, exhaustive analysis has been preferred instead of partial proofs or sampling.
Starting from the SA description CHARMY synthesizes, through a suitable translation into Promela (the specification language of the SPIN (Holzmann, 2003) model checker) a runnable SA prototype that can be executed and verified in SPIN. This model can be validated with respect to a set of properties expressed in the PSC language. By using CHARMY, thanks to a UML like notation used for the system design and the properties specification, we have an easy to use, practical approach to model and check architectural specifications, *hiding the modelling complexity*.
Whenever the SA specification does not properly implement selected requirements ("Not valid" arrow in Figure 4), the SA itself needs to be revised. Thanks to the model-checker outputs (i.e., a counter example reproducing the error) we may either correct the SA specification (if we discover that there is an error on the SA specification) or correct the PSC property (if we discover that the property is not properly expressed).

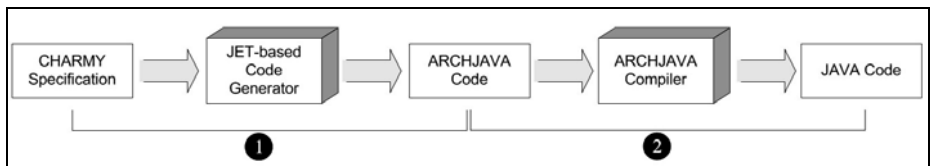
**Figure 6:** JET/ ARCHJAVA Code Generation

### 3.4 JET/ARCHJAVA Code Generation

Whenever the SA is validated with respect to the desired requirements, Java code is automatically generated from the SA specification. According to Figure 6, this activity is performed through two main steps: starting from a validated CHARMY *Specification,* ARCHJAVA *code* is automatically obtained by means of a *JET-based Code Generator*. Then, by exploiting the existing ARCHJAVA

*Compiler*, executable *Java code* is generated. Here we focus on the first step of the translation in Figure 6 which is based on the following directives:

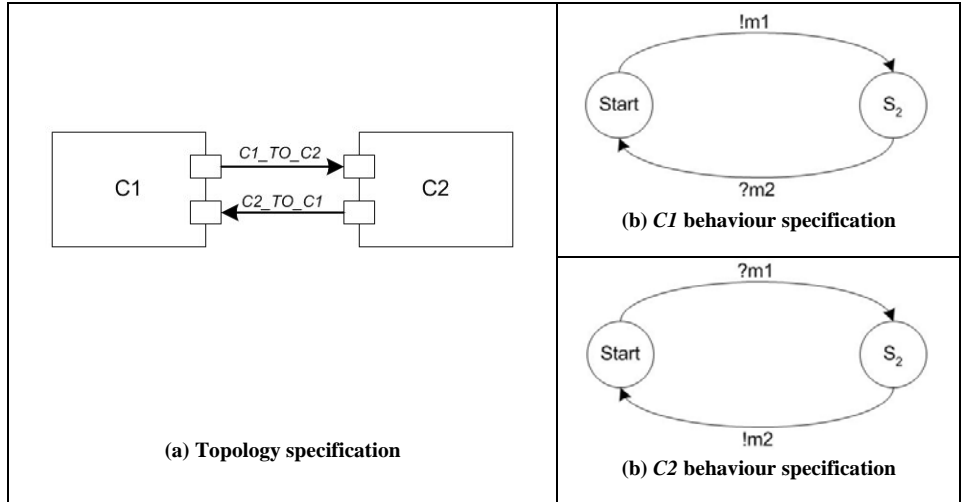

**Figure 7:** Sample CHARMY specification

**(i)** Each CHARMY component becomes an ARCHJAVA component. For instance, the component *C1* in Figure 7.a induces the following ARCHJAVA specification:

> **public** component **class** C1 {
> ...
> }

**(ii)** Each CHARMY component's sent and received message is used to synthesize the component ports. We recall that ARCHJAVA has both provided ports for provided services and required ports for required services. An ARCHJAVA port only connects a pair of components. This means that if a component needs to communicate with more than one component, it needs additional ports. Thus, the provided component services are partitioned into sets of services provided to different components. The same is done for required services. Accordingly, the suitable number of required and provided ports is declared into the ARCHJAVA specification of the component (containing the declaration of required and provided services, respectively). For instance, the sample SA in Figure 7.a gives place to the following ARCHJAVA code fragments concerning the *C1* component implementation:

> **public** port C1_TO_C2 {
>     requires **void** m1();
> }

```
public port C2_TO_C1 {
        provides void m2();
}
```

**(iii)** For each CHARMY component an ARCHJAVA specification is generated to encode the associated state diagram. ARCHJAVA does not offer a direct support for that and we propose guidelines to extend the ARCHJAVA specifications so that a state diagram associated to a software component is implemented as an adjacency list. In particular:
- for each method invoked by a given component the corresponding state diagram changes state accordingly so having trace of what methods can be invoked or not. States and transitions of the considered state diagram are declared as Java constants and are used to univocally refer to these elements (see lines 3-9 in Figure 8).
- each state machine contains a fixed definition of transitions as an internal Java class (see line 15-37 in Figure 8). The state diagram is defined as a *LinkedList*. The constructor of the state diagram class contains the definition of the state machine adding to the *LinkedList* of the state diagram an element for each state containing all existing transitions (for each existing transition a new object of the internal class transition is added) (see lines 40-50).
- each state diagram class contains also a method that simulates the transition fire, i.e., this method gets as input the transition (according to the runtime behaviour of the system) and checks if it is possible, in the actual computation state, to perform the transition fire (see lines 52-63). If the behaviour is allowed then the actual state is updated to the transition target state, otherwise an exception is raised. In case a method cannot be invoked in a certain time, an exception is raised. The exception is defined as an additional ARCHJAVA specification, i.e., a java class extending the *java.lang.Exception* class.

**(iv)** A *main* ARCHJAVA specification is also generated to define the binding among component's ports and the instantiation of the involved state machines

These directives ensure the *communication integrity*, i.e., components can only communicate using connections and ownership domains that are explicitly declared in the SA. The rest of the section outlines the approach supporting the automatic generation of code that implements such directives. This automation is required since manual coding could diverge or not completely adhere to them.

```java
1.    public class SM_C1 {
2.
3.    /** State encoding*/
4.    public final int S_startC1= 0;
5.    public final int S_S1= 1;
6.
7.    /** Transition encoding */
8.    public final int T_m1=0;
9.    public final int T_m2=1;
10.
11.   private int currentState=S_startC1;
12.
13.   private LinkedList states = new LinkedList();
14.
15.   private class transition{
16.           private int state;
17.           private int transition;
18.           private int send_receive;
19.
20.           public transition(int transition, int state, int send_receive){
21.                   this.transition=transition;
22.                   this.state=state;
23.                   this.send_receive=send_receive;
24.           }
25.
26.           public int getTransition(){
27.                   return transition;
28.           }
29.
30.           public int getState(){
31.                   return state;
32.           }
33.
34.           public int getSendReceive(){
35.                   return send_receive;
36.           }
37.   }
38.
39.   /** State Machine constructor*/
40.   public SM_C1(){
41.           System.out.println("SM_C1.constr");
42.
43.           LinkedList startC1 = new LinkedList();
44.           startC1.add(new transition(T_m1, S_S1 ,1));
45.           states.add(startC1);
46.
47.           LinkedList S_S1 = new LinkedList();
48.           S_S1.add(new transition(T_m2, S_startC1,0));
49.           states.add(S_S1);
50.   }
51.
52.   public void transFire(int trans) throws SMException {
53.           LinkedList transitions = (LinkedList) states.get(currentState);
54.           for (int i = 0; i < transitions.size(); i++) {
55.             if (((transition) transitions.get(i)).getTransition() == trans)
56.                currentState = ((transition) transitions.get(i)).getState();
57.                System.out.println("User.trans allowed: ");
58.                return;
59.             }
60.           }
61.           System.out.println("trans not allowed: " + trans);
62.           throw new SMException();
63.   }}
```

**Figure 8**: Sample state machine encoding

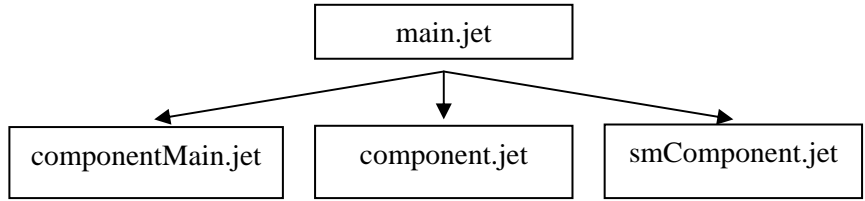

**Figure 9:** JET-based Code Generator templates

As previously said, the code generator implementing the four directives above has been developed in JET (Budinsky, 2003). JSP-like templates define explicitly the target ARCHJAVA code structure and get the data they need from the CHARMY models. In particular, the code generator consists of four templates: *main.jet* is a default template that gets data as input and applies the other templates. Being more precise, it applies the *componentMain.jet* template, that implements the directive *(iv)* previously described, producing the target *MAIN.archj* file (see line 2 in Figure 10). Then, for each component in the source CHARMY specification, the *component.jet* template is applied in order to generate the component implementation according to points *(i)* and *(ii)* above (see line 4-6 in Figure 10). Finally, for each source component the corresponding state machine encoding is generated by applying the *smComponent.jar* template that implements point *(iv)* (see line 8-10).

```
1.  ...
2.  <ws:file template="templates/componentMain.jet" path="{$org.eclipse.jet.resource.project.name}/src-
        generated/MAIN.archj"/>
3.
4.  <c:iterate select="$SAcomponent" var="component">
5.  <ws:file template="templates/component.jet" path="{$org.eclipse.jet.resource.project.name}/src-
        generated/{$component/@nome}.archj"/>
6.  </c:iterate>
7.
8.  <c:iterate select="$SAcomponent" var="component">
9.  <ws:file template="templates/smComponent.jet" path="{$org.eclipse.jet.resource.project.name}/src-
        generated/SM_{$component/@nome}.archj"/>
10. </c:iterate>
```

**Figure 10:** Fragment of the main.jet tamplate

Due to space limitation, the templates are not reported here. However, interested readers can refer to (CHARMY, 2004) for downloading the full implementation of the proposed JET-based code generator.

In the next section we will apply the architecture-based approach we are proposing on a case study. We will focus principally on the code generation phase, to be considered the main contribution of this chapter.

## *4. Explanatory example: an ATM System*

A bank has several automated teller machines (ATMs), which are geographically distributed and connected to a central server. Each ATM has a card reader, a cash dispenser, a keyboard/display, and a receipt printer. An user can withdraw cash or recharge a mobile phone credit. Assuming that the card is recognised, the system validates the ATM card to determine that the card has not expired and that the user-entered PIN (Personal Identification Number) is correct. If the user is authorized, it is prompted for withdraw or recharge transaction. Before these transactions can be approved, the bank determines that sufficient funds exist in the requested account. If the transaction is approved, the requested amount of cash is dispensed, the account is updated and the card is ejected. An user may cancel a transaction at any time with a logout operation.

### 4.1. Functional Requirements Specification

In this section we report only a subset of requirements useful for explaining the approach. The requirements are presented in the following as use case tables. The first one, in Table 4, is the User login use case that describes the user interactions to get access to the ATM.

| Use Case Name | User login |
|---|---|
| Description | The ATM System validates the USER PIN |
| Actors | USER, TM, AUTH |
| Pre-Conditions | ATM is idle, displaying a Welcome message |
| Process Steps | 1. USER enters the PIN (*login*). <br><br> 2. TM forwards the request (*login_auth*) to AUTH that checks whether the USER-entered PIN matches the card PIN maintained by the system. <br><br> 3. If PIN numbers match, AUTH notifies it (*login_auth_ok*) to TM. <br><br> 4. TM notifies to USER (*login_ok*) the login successful and prompts customer for transactions type (*withdraw* or *chargePhone*). |
| Post-Conditions | USER has been validated |
| Alternative Paths | If the USER-entered PIN does not match the PIN number of the card, AUTH notifies to TM an error (*login_auth_ko*) and TM asks USER to re-insert the PIN (*login_ko*). |

**Table 4:** User login use case

The other use case, represented in Table 5, is the withdraw functionality of the ATM that allows the user to withdraw money from the bank. This use case includes the Use login use case as precondition.

| Use Case Name | Withdraw |
|---|---|
| Description | USER withdraws a specific amount of money from a valid bank account |
| Dependency | Include Validate PIN Use Case |
| Actors | USER, TM, AUTH, BA |
| Pre-Conditions | ATM is idle, displaying a Welcome message |
| Process Steps | 1. Include User login use case<br>2. USER selects *withdraw* and enters the amount of money to be withdrawn<br>3. TM forwards the request to BA (*connect*)<br>4. If the request is accepted BA notifies the connection to TM (*connect_ok*)<br>5. TM checks whether USER has enough money by BA (*check_funding*)<br>6. If USER has enough money BA notifies it to TM (*funding_ok*)<br>7. TM dispenses the cash amount (*withdraw_ok*)<br>8. USER gets the amount of money and the card (*logout*) |
| Post-Conditions | USER money have been withdrawn |
| Alternative Paths | • If TM experiences problems that can compromise the operation, it sends an error to BA (*noconnection*) and the TM ejects the card (*logout*).<br>• If the BA determines that there are insufficient funds in the USER's account, it notifies it to TM (*funding_ko*) and the TM ejects the card (*logout*). |

**Table 5:** Withdraw use case

### 4.2. Software Architecture Specification in CHARMY

The architecture of the ATM system that we consider is composed of four components as shown in Figure 11: the user (USER component), the transaction manager (TM component), the bank account (BA component), and the authentication manager (AUTH component). The USER component communicates only with the TM component that forwards the service requests to the BA component or to the AUTH component.

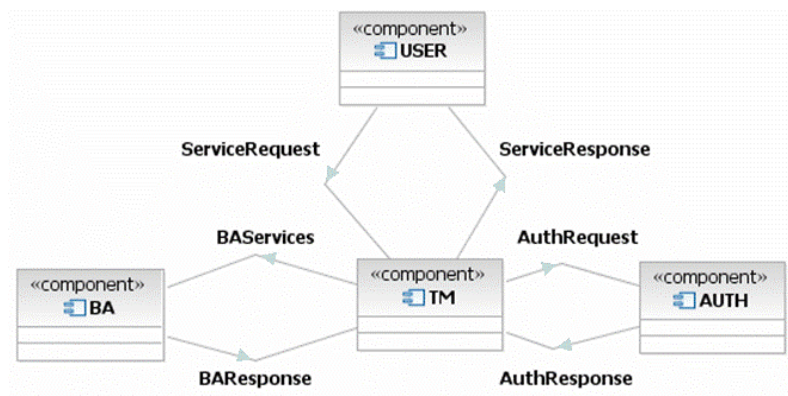

**Figure 11:** Software Architecture of the ATM System

Moreover the behaviour of each component is described with the state machines depicted in Figures 12-14. The USER component (refer to Figure 12) handles three different requests, one for the authentication (*!login*) followed by two possible responses (*?login_ok* and *?login_ko*), one for withdrawing money from its account (*!withdraw*), and one for recharging the mobile phone credit (*!chargePhone*).

The TM component (refer to Figure 14) contains the logic of the ATM system. This component receives the login request from the User (*?login*) and forwards it to the AUTH component (*!login Auth*).

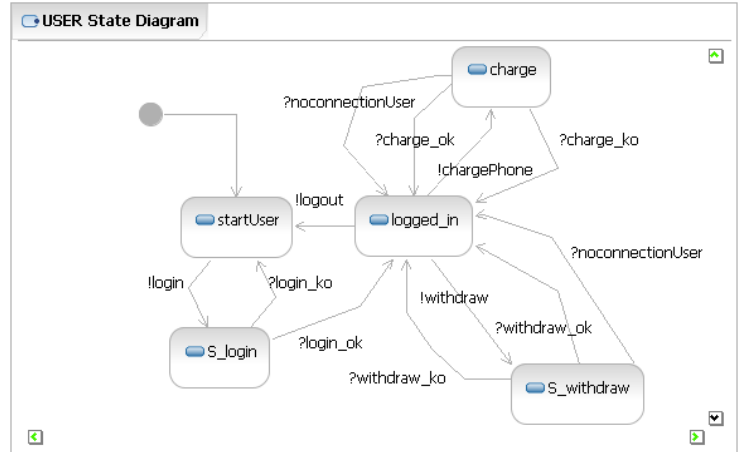

**Figure 12:** USER state diagram

Two are the possible responses that TM can receive from AUTH: login success (*?login_auth_ok*), and login failure (*?login_auth_ko*). The state diagram of the AUTH component is shown in Figure 13, left-hand side. In case of success, the user is habilitated to available services (i.e., withdraw money or recharge mobile phone). TM receives the response for both services and forwards them to the User component. The other component, BA (right-hand side Figure 13), manages the bank account services (i.e., withdraw, charge).

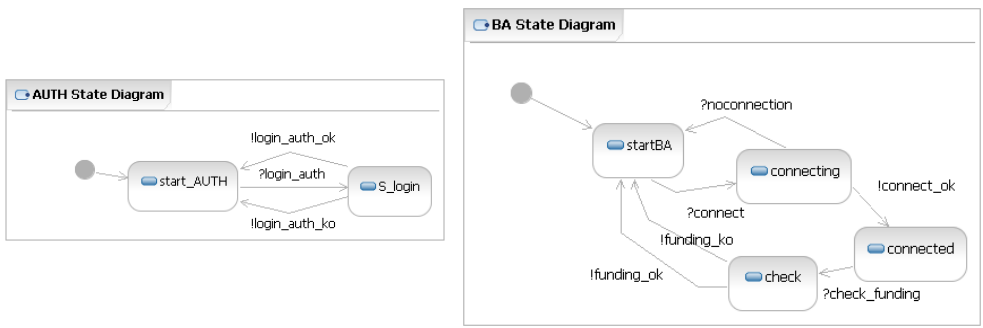

**Figure 13:** AUTH and BA state diagrams

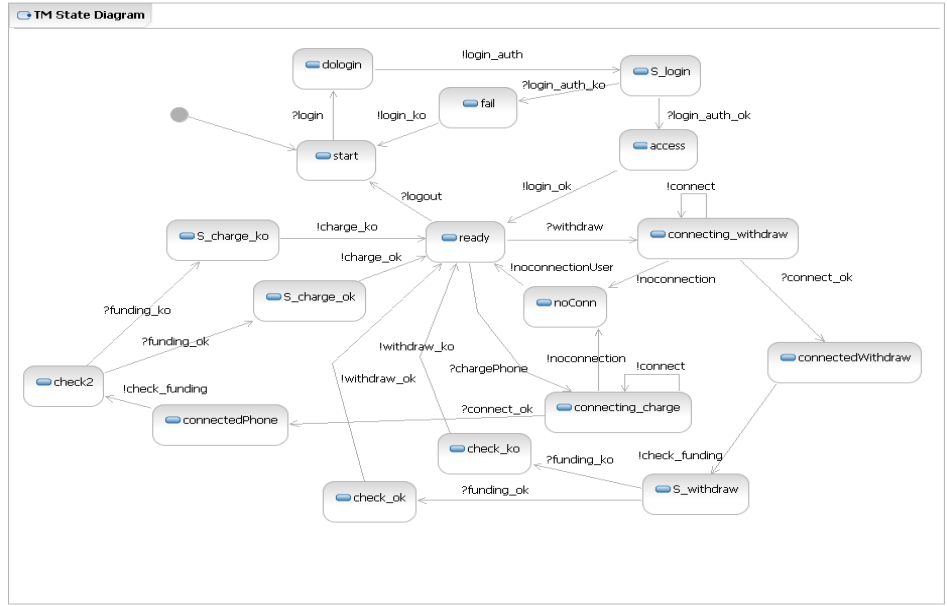

**Figure 14:** TM state diagram

## 4.3. W_PSC Requirements Specification and formalization in the PSC language

Starting from the two use cases selected in the previous subsection, a set of properties to be checked on the system are extracted. In the following we will explain two properties and we will provide their descriptions as PSC. The formalization of the properties as PSC is made by using W_PSC.

**Property 1**: if the withdraw request has been performed (*withdraw*) and before no errors on the connection have been raised (*noconnection* is not sent), and the request of money is consistent with the user funds (*funding_ok*), then TM must dispense the cash amount (*withdraw_ok*); the withdraw request is allowed only after a successful login request.

Having requirements well formulated as the ones considered in this chapter is an ideal situation not very common in real projects. W_PSC can be particularly useful in these situations since, as already explained, it is not an automatic tool but it is a wizard that helps the designer in making decisions in formalizing the requirements and in restructuring them with the required accuracy. In the following we provide enough information on W_PSC and PSC for understanding the case study and we refer to (Autili, 2006; Autili, 2007) for further details.

An important aspect to be considered when formalizing the requirements is the distinction among *Mandatory*, *Forbidden* and *Optional* operations. They are organized into W_PSC as different sentences contained into 3 different panels. Thus, reading Property 1 the first action that can be found is *login*. It is easy for the software engineer to understand if the considered part of the requirement is mandatory, forbidden or optional. Making this decision the suitable tab panel containing the pre-formulated sentences is chosen. *login* is clearly an optional operation since the exchange of the message *login* represents the precondition for *withdraw* and for the following messages. In the Optional panel among the proposed sentences (and reported in the following), the software engineer selects the Sentence 1 since no other constraints on the login message are required.

**Sentence 1** If the message *< m >* is exchanged then ...
**Sentence 2** If the message *< m >* is exchanged and between this message and its predecessor (or the system startup) no other messages can be exchanged then ...
**Sentence 3** If the message *< m >* is exchanged and between this message and its predecessor (or the system startup) *< … >* then ...
**Sentence 4** If the message *< m >* is exchanged and between this message and its successor (or after the last message) *< … >* then ...
**Sentence 5** If the message *< m >* is exchanged and between this message and its predecessor (or the system startup) *< … >* and between this message and its successor (or after the last message) *< … >* then ...

Following a similar reasoning *withdraw* is identified as another optional message but in this case the selected sentence is the Sentence 3 since *withdraw* is a valid precondition if and only if before this message no connection errors have been raised. *Funding* is another precondition, in this case without constraints, for the final message of the property that is *withdraw_ok*. *withdraw_ok* is a mandatory message (i.e., the correct sentence will be selected among the Mandatory sentences) since the system is in error if the this message is not exchanged.

When the formalization is finished in W_PSC the corresponding PSC is automatically generated.

The PSC corresponding to Property 1 is depicted in Figure 15. In PSC messages are typed and, in particular, messages prefixed by **"e:"** denote messages not mandatory for the system and are used for constructing the precondition for mandatory ones. Mandatory messages on the contrary are prefixed by **"r:"** and denote messages that must be exchanged by the system, i.e., if the messages are not exchanged then the system is in error. The circle labelled **b** identifies a constraint of the message. It means that the message *withdraw* that has associated the constraint is a valid precondition iff before this message and after the previous one (i.e., *login*) TM does not send *noconnection* to BA.

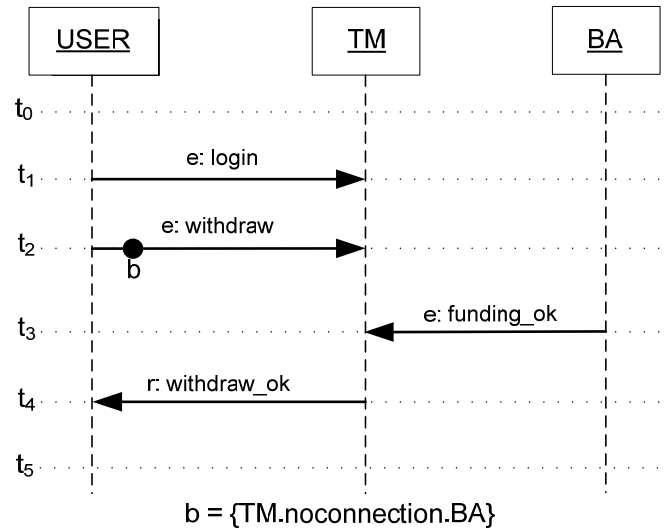

**Figure 15:** PSC of Property 1

Thus, Property 1 expresses that if USER sends *login* to TM and after that it sends the *withdraw* request to TM, if in between of these two messages *noconnection* has been not sent by TM to BA, and if BA sends *funding_ok* to TM, then TM **must** send *withdraw_ok* to USER (i.e., the user receives the requested money).

**Property 2**: if the withdraw request has been performed (*withdraw*), there are no errors (*noconnection* is not sent), and the request of money is not consistent with the user funds (*funding_ko*), then there is an error if TM dispenses the cash amount (*withdraw_ok*); the withdraw request is allowed only after a successful login request.

The PSC generated by W_PSC for Property 2 is the one in Figure 16. In PSC messages can be also typed as fail, i.e., prefixed by **"f:"**, i.e., messages used to model erroneous behaviours of the system.

Thus, Property 2 expresses that if USER sends *login* to TM and after that it sends the *withdraw* request to TM, if in between of these two messages *noconnection* has been not sent by TM to BA, and if BA sends *funding_ko* to TM, then if TM sends *withdraw_ok* to USER the system is in error (i.e., the user cannot receive the requested money).

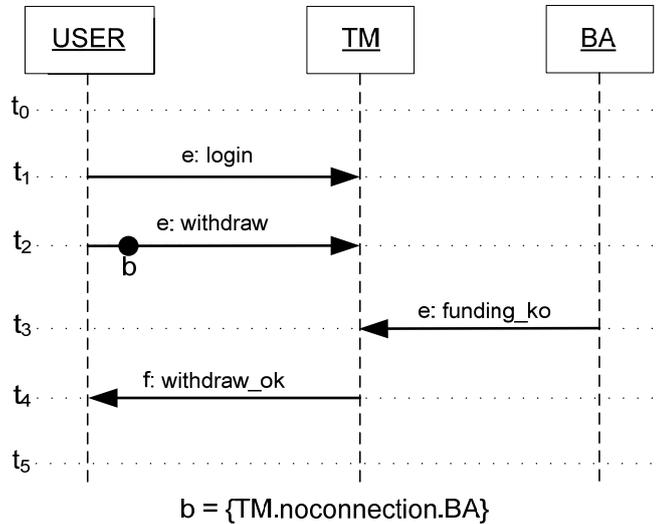

**Figure 16:** PSC of Property 2

### 4.4. Software Architecture Verification

The ATM software architecture presented in the previous section has been modelled in CHARMY, and the verification has been performed.

The first verification concerns the deadlocks detection. The verification is performed on a Pentium 1.73Ghz with 1,50 GB of RAM and took less that 1 minute using 2,582 MB of memory. The system specification is deadlock free and has 311 states and 663 transitions. Furthermore, there are no unreachable parts of the model.

The next step is the verification of the properties.

**Property 1:** this property is valid. The number of generated states is 1535, while the transitions are 33724. The memory used in this case is 2.622 MB of RAM.

**Property 2:** this property is also valid. The number of generated states is 951, with 2046 transitions. The memory used is 2.582 MB of RAM.

A. Bucchiarone, D. Di Ruscio, H. Muccini, P. Pelliccione, *From Requirements to Java code: an Architecture-centric Approach for producing quality systems*

Now the software architecture is verified and then it can be used as starting point for the implementation, as real blueprint for the development. The next section shows the code generation phase and shows how software architecture choices force the implementation.

### 4.5. JET/ARCHJAVA code generation

The application of the *JET-based Code Generator* (outlined in Section 3.2) on the CHARMY specification of the ATM case study produces a number of ARCHJAVA files listed on the left-hand side of the screenshot in Figure 17. In particular, for each component (e.g., *User*), the corresponding encoding is generated (e.g., *User.archj*). The state machine specifications are also synthesized (e.g., *SM_User.archj*) together with a *MAIN.archj* file (listed on the right-hand side of Figure 17) that enables the execution of the obtained system with respect to the modelled software architecture. Being more precise, in that main file all the components, the corresponding state machines and port connections are instantiated giving place to an encoding of the SA properties that constraint the execution of the hand-written code that will be filled in prearranged points (e.g., see the *try* statement in the code of Figure 17).

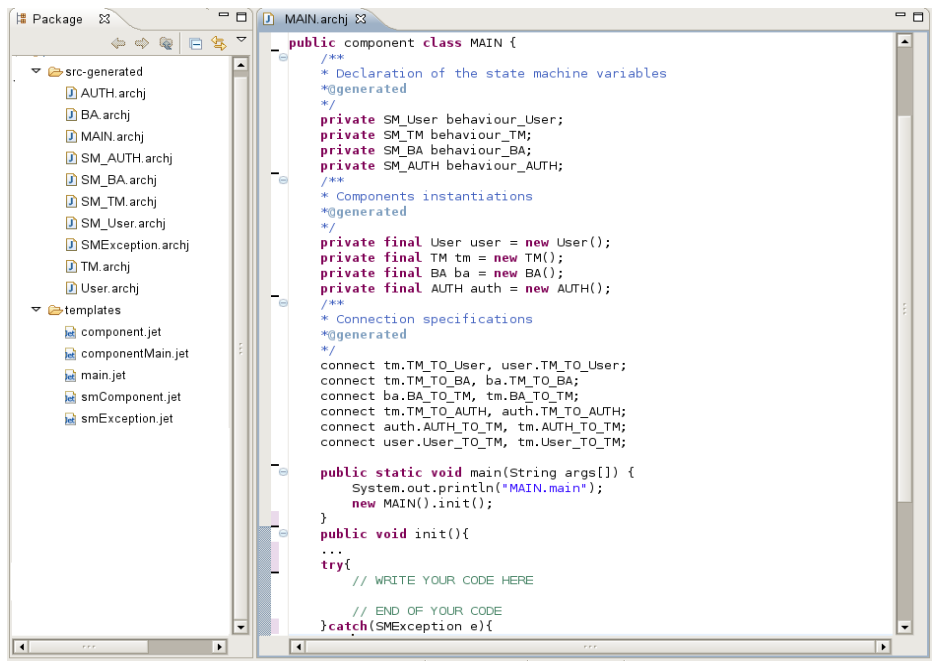

**Figure 17: Generated Code Overview**

```
1.    public component class User {
2.    /**Declaration of the state machine variables
3.     *@generated
4.     */
5.    private SM_User behaviour_User;
6.    private SM_TM behaviour_TM;
7.    private SM_BA behaviour_BA;
8.    private SM_AUTH behaviour_AUTH;
9.    /**TM_TO_User Port definition
10.    *@generated
11.    */
12.   public port TM_TO_User {
13.        provides void login_ko() throws SMException;
14.        provides void login_ok() throws SMException;
15.        provides void charge_ko() throws SMException;
16.        provides void charge_ok() throws SMException;
17.        provides void withdraw_ko() throws SMException;
18.        provides void withdraw_ok() throws SMException;
19.        provides void noconnectionUser() throws SMException;
20.   }
21.   /**User_TO_TM Port definition
22.    *@generated
23.    */
24.   public port User_TO_TM {
25.        requires void withdraw() throws SMException;
26.        requires void chargePhone() throws SMException;
27.        requires void login() throws SMException;
28.        requires void logout() throws SMException;
29.   }
30.   /**Implementation of the methods provided by the port TM_TO_User
31.    *@generated
32.    */
33.   public void login_ok() throws SMException {
34.        System.out.println("User.login_ok");
35.        behaviour_User.transFire(behaviour_User.T_login_ok);
36.        behaviour_TM.transFire(behaviour_TM.T_login_ok);
37.        //WRITE YOUR CODE HERE
38.        //END YOUR CODE HERE
39.   }
40.   public void withdraw_ko() throws SMException {
41.        System.out.println("User.withdraw_ko");
42.        behaviour_User.transFire(behaviour_User.T_withdraw_ko);
43.        behaviour_TM.transFire(behaviour_TM.T_withdraw_ko);
44.        //WRITE YOUR CODE HERE
45.        //END YOUR CODE HERE
46.   }
47.   public void withdraw_ok() throws SMException {
48.        System.out.println("User.withdraw_ok");
49.        behaviour_User.transFire(behaviour_User.T_withdraw_ok);
50.        behaviour_TM.transFire(behaviour_TM.T_withdraw_ok);
51.        //WRITE YOUR CODE HERE
52.        //END YOUR CODE HERE
53.   }
54.   public void login_ko() throws SMException {
55.        System.out.println("User.login_ko");
56.        behaviour_User.transFire(behaviour_User.T_login_ko);
57.        behaviour_TM.transFire(behaviour_TM.T_login_ko);
58.        //WRITE YOUR CODE HERE
59.        //END YOUR CODE HERE
60.   }
61.   ...
62.   }
```

**Figure 18:** Fragment of the generated User.archj

Focusing on the User component, a fragment of the corresponding generated code is listed in Figure 18. Essentially, it contains the declaration of the state machines that will be considered during the User component execution (hence the state machine of the User, TM, BA and AUTH components) (see lines 5-8 in Figure 18), the definition of the ports (*TM_TO_User* and *User_TO_TM,* lines 12-29) and the implementation of the *provided* methods that have to be completed by the developer (lines 33-59). The generated method statements are devoted to fire the transitions of the involved state machines. For instance, the *withdraw_ko()* invoked by the *TM* component induces the state changes in the *User* and *TM* state machines. The former will reach the state *S_withdraw* from the *logged_in* one, whereas the latter change the state *check_ko* reaching the *ready* one, according to the state machines in Figure 12 and Figure 14.

In order to have a full application, the developer has to complete the generated code by implementing the logic of each *provided* method. The hand-written code can be filled in the predefined user regions like the one in lines 44-45. The code that will be written in such blocks will never be updated by subsequent generations. This facility is provided by the JET framework and it is a first step towards round-trip engineering and refactoring support even though a more advanced support is required and it is an issue that has to be deeply investigated in the future.

The generated specification of the *User* state machine that guarantees the code execution according to the admitted *User* component behaviour is listed in Figure 19. The states and the transitions are encoded as a linked list initialized in the lines 32-57 of the Figure. This specification forces the execution of the methods which are admitted in the current computation state with respect to both SA constraints and behavioural one: if a given *User* method can be invoked, then the current state is updated according to the information contained in the linked list. Otherwise an exception is raised. For instance, once the *User* component reach the *logged_in* state, only the *withdraw, chargePhone,* and *logout* transitions are admitted (see lines 41-45 in Figure 19), consistently with the *User* state diagram (see Figure 12). If a different transition is asked, an exception is raised stopping the system execution.

A. Bucchiarone, D. Di Ruscio, H. Muccini, P. Pelliccione, *From Requirements to Java code: an Architecture-centric Approach for producing quality systems*```
1.     ** User State Machine encoding
2.     *@generated
3.     */
4.     public class SM_User {
5.     /** State encoding
6.     *@generated
7.     */
8.     public final int S_startUser= 0;
9.     public final int S_S_login= 1;
10.    public final int S_logged_in= 2;
11.    public final int S_S_withdraw= 3;
12.    public final int S_charge= 4;
13.    /** Transition encoding
14.    *@generated
15.    */
16.    public final int T_login_ok=0;
17.    public final int T_withdraw=1;
18.    public final int T_chargePhone=2;
19.    public final int T_charge_ko=3;
20.    public final int T_withdraw_ko=4;
21.    public final int T_withdraw_ok=5;
22.    public final int T_login=6;
23.    public final int T_login_ko=7;
24.    public final int T_logout=8;
25.    public final int T_charge_ok=9;
26.    public final int T_noconnectionUser=10;
27.    /** State Machine constructor
28.    *@generated
29.    */
30.    public SM_User(){
31.        System.out.println("SM_User.constr");
32.        LinkedList startUser = new LinkedList();
33.        startUser.add(new transition(T_login, S_S_login ,0));
34.        states.add(startUser);
35.
36.        LinkedList S_login = new LinkedList();
37.        S_login.add(new transition(T_login_ok, S_logged_in ,1));
38.        S_login.add(new transition(T_login_ko, S_startUser ,1));
39.        states.add(S_login);
40.
41.        LinkedList logged_in = new LinkedList();
42.        logged_in.add(new transition(T_withdraw, S_S_withdraw ,0));
43.        logged_in.add(new transition(T_chargePhone, S_charge ,0));
44.        logged_in.add(new transition(T_logout, S_startUser ,0));
45.        states.add(logged_in);
46.
47.        LinkedList S_withdraw = new LinkedList();
48.        S_withdraw.add(new transition(T_withdraw_ko, S_logged_in ,1));
49.        S_withdraw.add(new transition(T_withdraw_ok, S_logged_in ,1));
50.        S_withdraw.add(new transition(T_noconnectionUser, S_logged_in ,1));
51.        states.add(S_withdraw);
52.
53.        LinkedList charge = new LinkedList();
54.        charge.add(new transition(T_charge_ko, S_logged_in ,1));
55.        charge.add(new transition(T_charge_ok, S_logged_in ,1));
56.        charge.add(new transition(T_noconnectionUser, S_logged_in ,1));
57.        states.add(charge);
58.    }
```

**Figure 19:** Fragment of the generated SM_User.archj





## 5. *Future Trends*

In this chapter we presented an approach to automatically generate the code starting from verified software architecture descriptions. The best of the state of the art, as presented in Section 2, is represented by ARCHJAVA that ensures the communication integrity, and by JAVA/A that constraints also the code to behave as defined in the port's protocols. These approaches (as they are today) can be used only in a context in which the system is completely implemented in-house, while neglecting the possibility of integrating external components. This is because acquired components are not necessarily implemented following one of these approaches and thus it is not possible at runtime to allow the only admitted operations. Thus, an interesting future research direction consists of the ability of integrating in-house components code with automatically generated assembly code for acquired components, forcing the composed system to exhibit the properties specified at the architectural level. This integration would open the possibility of managing dynamic systems (i.e., systems in which some components have to change at runtime) where a re-generated "correct" assembly code assures that the composed system is forced to exhibit only the properties specified at the architectural level.

In the domain of system run-time validation, feedback generated by the run-time analysis of the generated code (through monitoring and testing techniques) could be automatically tracked back to the architectural model, so that whenever a change applies over the code, it is automatically reflected on the architectural model and vice-versa.

Another future research direction consists in investigating how to assure that the generated code respects non functional properties and quality aspects which have been proven to be valid at the architectural level.

## 5. *Related Work*

The principal aspect related to our work is the use of "Software Architecture" in a development process of complex and distributed systems.

Although the concept of "Software Architecture" was first described in the 60s, a significant interest among the research community is only 15 years old, and the industrial community's interest is very recent. Software Architecture's research results are widely ignored by the industrial community for which a good architect is mainly an experienced person. Currently the most used technique for distributed systems engineering is based on "best practices" documents that do not necessarily include precise models and descriptions.

There are few development methodologies covering architecture engineering. Some interesting initiatives covering architecture engineering are RUP (RUP 2000), MDA (OMG), HP (HP 1998) recommendations initially started in the Fusion 2.0 Project and the FIDJI project (FIDJI 2001).





In RUP, UML models should be used to describe all the architecture artefacts. The famous "4+1" views (Kruchten, 1995) have been introduced to capture heterogeneous architectural properties that has to be understood by many people who have various jobs and therefore various background.

In the OMG Model Driven Architecture (MDA) the architecture is at the center of software development. The key concepts are: *PIM* (Platform Independent Model) which is an abstract description of the system being developed. In particular, a PIM exhibits a specified degree of platform independence so as to be suitable for use with a number of different technologies; *PSM* (Platform Specific Model) which is a view of a system from the platform specific viewpoint. A PSM combines the specifications in the PIM with architectural details and specifies how that system uses a particular platform. The approach proposed in this chapter adheres to the MDA ideas even though it mainly highlights architectural aspects of software systems.

Fusion 2.0 (HP 1998) defines an architecture phase placed between analysis and design. It is split in "conceptual architecture" and "logical architecture"; the first one is more informal and abstract and the second one is very precise. This notion is very close to the PIM and PSM in MDA.

The intent of the FIDJI project is to define a methodology for distributed applications in Java. The approach is composed on four steps that are inspired from Fusion 2.0: *Requirements*, *Domain analysis*, *Architectural* and *Design*.

In the requirements step it is possible to define a set of use-cases inter use-cases relationships and contracts. Contracts are statements that have to be satisfied by the final system. The domain analysis step provides criteria to choose API and components that already exist. From domain analysis and requirements this approach is able to identify the architectural style that will act as guide for the remaining design. From the style and using a *framework specialization* tool (UML, ADL, etc…) the architecture is defined, developed and stabilized. The most difficult issue of this approach is to provide a concrete architecture that support the functional requirements and which validates its associated non-functional requirements.

Overall, the main distinction between our and existing approaches is that while they cover some of the steps in the software development process, we do cover the entire process from requirements to code. We specifically focus on the Software Architecture as the main artefact enabling the transition from





requirements to code. Moreover, we especially focus on quality during the entire process.

## 6. Conclusions

Model-driven development is based on the idea that code can be generated from a model of the system. This chapter has provided a contribution in this direction, by showing how an architectural model defined in a model-based fashion can be used for code generation. An important aspect during this process consists in ensuring that the selected architecture provides the required qualities. We have shown how this is feasible in a specific context, where coordination properties are elicited from requirements, modelled, and verified against the architectural model. As soon the architectural model is proven to be good enough, we demonstrated that it can be used for generating Java code, constraining the system execution according to the architectural decisions.

Indeed, this approach shall be considered as a mere feasibility study, which demonstrates one possible way of achieving the desired objectives, while motivating other researchers to pursue similar objectives.

### Acknowledgment

The authors of this chapter wish to thank the anonymous reviewers for their careful and useful comments. This work has been partially supported by the Italian FIRB Project ART DECO (Adaptive InfRasTructures for DECentralized Organizations).

## Appendix A: JAVA/A CODE GENERATION

The following code shows parts of the Java/A declaration of the component *OutService* described in Figure 3:

```
1.  simple component AccidentAssistanceService {
2.  dynamic port AAE {
3.     provided {
              void AlertAccepted();
              void AlertNotAccepted();
              void EmergencyAccepted();
              void EmergenctNotAccepted();
         }
4.     required {
              signal AlertEmergencyService(Location Loc);
              signal EmergencyLevel(int Level);
              void AlertAccepted();
         }
5.
6.  try{
7.         Component aas = componentLookUp (this,
                  "AccidentAssistenceService");
8.         Port aae = aas.getPort ("AAE");
9.         ConnectionRequest cr = (this, this, EAA, aas, aae, new
                  Connector());
           reconfigurationRequest(cr);
10. }
11. Catch (ReconfigurationException e) {…}
12. }
13. simple component EmergencyService{
14. port EAA {
15. provided{
              signal AlertEmergencyService(Location Loc);
              signal EmergencyLevel(int Level);
              void AlertAccepted();
         }
16. required{
              void AlertAccepted();
              void AlertNotAccepted();
              void EmergencyAccepted();
              void EmergenctNotAccepted();
17. }
18. <! // protocol of EAA
         states {
                initial Initial;
                simple Q1,Q2,Q3,Q4;
              }
         transitions {
           Initial -> Q1;
           Q1 -> Q2 {trigger AlertEmergencyService();}
           Q2 -> Q1 {effect AlertNotAccepted();}
           Q2 -> Q3 {effect AlertAccepted();}
           Q3 -> Q4 {trigger EmergencyLevel();}
           Q4 -> Q3 {effect EmergencyNotAccepted();}
           Q4 -> Q1 {effect EmergencyAccepted();}
```



A. Bucchiarone, D. Di Ruscio, H. Muccini, P. Pelliccione, *From Requirements to Java code: an Architecture-centric Approach for producing quality systems*

```
                }
        !>

19. }
20. composite component OutService
21. {
22.     assembly {
            component types {AccidentAssistenceService,
            EmergencyService
            }
23. connector types {
        AccidentAssistenceService.AAE;
        EmergencyService.EAA;
                }
24. initial configuration {
       AccidentAssistenceService AS = new AccidentAssistenceService();
       EmergencyService ambulance = new EmergencyService();
        EmergencyService police = new EmergencyService();
         Connector cn0 = new Connector();
         cn0.connect(ambulance.EAA, AS.AAE);
          connector cn1 = new Connector();
          cn1.connect = (police.EAA, AS.AAE);
                }
         }
      }
25. }
```

In lines 1-12 and 13-19 the two simple components (Accident Assistence Service (AAS) and Emergency Service (ES)) are declared while in lines 20-25 a composite component "*OutService*" is declared as an assembly of the two previous ones. In lines 2 and 14, the ports AAE and EAA are defined. Each port declaration contains a set of provided operations (i.e., lines 3 and 15) and a set of required operations (i.e., lines 4 and 16). Port protocols are specified by UML state machines which are textually represented using the notation UTE (HUGO, 2005). For instance, lines 18-19 show the UTE representation of the UML state machine for the port EAA. In line 24 a possible configuration of the *OutService* composite component is declared. It presents two instances of the ES component (ambulance and police) that are attached at the AAS by the EAA and AAE ports. The last interesting thing the JAVA/A can models is the "*dynamic reconfiguration*" that describes changes to a component-based system at run-time, concerning creation and destruction of components and building up and removing connections between ports. This is made with a code like to lines 6-11 where a possible reconfiguration in the *OutService* composite component (i.e., the connection and disconnection of ES) is presented. An idle ES disconnects from the AAS and reconnects whenever there is an accident and the AAS alerts the ES. When AAS alerts the ES executes the code in the 6-11 lines which realised the (re)connection of an ES to the AAS.